\documentclass[twocolumn,showpacs,preprintnumbers,amsmath,amssymb,10pt,prd]{revtex4-1}
\usepackage{fancyhdr}
\usepackage{amsmath,amsfonts,amssymb}
\usepackage{graphicx}
\usepackage{dcolumn} 
\usepackage{bm}
\usepackage{hyperref}
\usepackage[pdftex]{color}
\usepackage{mathrsfs}
\usepackage{mathtools}
\usepackage{float}
\usepackage{color}
\definecolor{red}{rgb}{1,0,0}
\definecolor{gre}{rgb}{0,0.6,0}
\definecolor{blu}{rgb}{0,0,1}
\usepackage{graphicx} 
\usepackage{mathrsfs}
\usepackage{ulem}
\DeclareGraphicsExtensions{.jpg, .JPG , .png , .pdf, .gif}

\def\Hc{\mathscr{H}}

\newcommand{\bark}{\bar{k}}
\newcommand{\barp}{\bar{p}}

\newcommand{\barn}{\bar{N}}

\newcommand\be{\begin{equation}}
\newcommand\ee{\end{equation}}
\newcommand\ba{\begin{eqnarray}}
\newcommand\ea{\end{eqnarray}}

\newcommand{\qsubrm}[2]{{#1}_{\scriptscriptstyle{\textrm{#2}}}}

\DeclarePairedDelimiter\ket{\lvert}{\rangle}
\DeclarePairedDelimiterX\braket[2]{\langle}{\rangle}{#1 \delimsize\vert #2}

\begin{document}
\title{Primordial scalar power spectrum from the Euclidean Big Bounce}
\author{Susanne Schander}%
 \email{schander@lpsc.in2p3.fr}
\affiliation{%
Technische Universit\"at Kaiserslautern, D-67653 Kaiserslautern, Germany
}%
\affiliation{%
Laboratoire de Physique Subatomique et de Cosmologie, Universit\'e Grenoble-Alpes, CNRS-IN2P3\\
53, avenue des Martyrs, 38026 Grenoble cedex, France
}%
\author{Aur\'elien Barrau}%
 \email{Aurelien.Barrau@cern.ch}
\affiliation{%
Laboratoire de Physique Subatomique et de Cosmologie, Universit\'e Grenoble-Alpes, CNRS-IN2P3\\
53, avenue des Martyrs, 38026 Grenoble cedex, France
}%
\author{Boris Bolliet}%
 \email{boris.bolliet@ens-lyon.fr}
\affiliation{%
Laboratoire de Physique Subatomique et de Cosmologie, Universit\'e Grenoble-Alpes, CNRS-IN2P3\\
53, avenue des Martyrs, 38026 Grenoble cedex, France
}%
\author{Linda Linsefors}%
 \email{linsefors@lpsc.in2p3.fr}
\affiliation{%
Laboratoire de Physique Subatomique et de Cosmologie, Universit\'e Grenoble-Alpes, CNRS-IN2P3\\
53, avenue des Martyrs, 38026 Grenoble cedex, France
}%
\author{Jakub Mielczarek}%
 \email{jakub.mielczarek@uj.edu.pl}
\affiliation{%
Laboratoire de Physique Subatomique et de Cosmologie, Universit\'e Grenoble-Alpes, CNRS-IN2P3\\
53, avenue des Martyrs, 38026 Grenoble cedex, France
}%
\author{Julien Grain}%
 \email{julien.grain@ias.u-psud.fr}
\affiliation{%
CNRS, Orsay, France, F-91405
}%
\affiliation{%
Universit\'e Paris-Sud 11, Institut d'Astrophysique Spatiale, UMR8617, Orsay, France, F-91405
}%
\date{\today}
\begin{abstract}
In effective models of loop quantum cosmology, the holonomy corrections are associated with deformations of space-time symmetries. The most evident manifestation of the deformations is the emergence of an Euclidean phase accompanying the non-singular bouncing dynamics of the scale factor. In this article, we compute the power spectrum of scalar perturbations generated in this model, with a massive scalar field 
as the matter content.  Instantaneous and adiabatic vacuum-type initial conditions for scalar perturbations are imposed in the contracting phase. The evolution 
through the Euclidean region is calculated based on the extrapolation of the time direction pointed by the vectors normal to the Cauchy hypersurface in the Lorentzian domains. The obtained power spectrum is characterized by a suppression in the IR regime and oscillations in the intermediate energy range. Furthermore, the speculative extension of the analysis in the UV reveals a specific rise of the power leading to results incompatible with data.
\end{abstract}
\maketitle
\section{Introduction}
Loop quantum gravity (LQG) is a simple, consistent, non-perturbative and background-independent 
quantization of general relativity. It uses Ashtekar variables, namely the SU(2)-valued connections and 
the conjugate densitized triads. The quantization is obtained through holonomies of the connections 
and fluxes of the densitized triads. No heavy hypothesis is required. Introductions can be found in Refs.~\cite{lqg_review}. \\
Loop quantum cosmology (LQC) is an application of  LQG-inspired quantization methods to a gravitational system with cosmological symmetries. In LQC, the big bang is generically replaced by a big bounce due to repulsive quantum geometrical effects when the density approaches the Planck density and interesting predictions can be made about the duration of inflation when a given matter content is assumed. It is, however, important to underline that LQC has not yet been rigorously derived from LQG and remains an attempt to use LQG-like methods in the cosmological sector. 
Introductions can be found in Refs.~\cite{lqc_review,Ashtekar_Review}.\\
The confrontation of LQG with available empirical data is crucial in order to check the physical validity of this approach to quantum gravity. The most promising option in this direction is currently given by the exploration of the cosmological sector of LQG. The present state of advancement, however, does not allow for a derivation of the cosmological dynamics directly from the full theory. Because of this, LQC models are considered to fill the existing gap. These models suffer from quantum ambiguities, which are believed to be 
fixed by the cosmological dynamics regained from LQG.\\

This study is based on the effective LQC dynamics, which allow to address various cosmological issues.
In particular, numerous studies have been devoted to the computation of tensor power spectra and their 
significance in the light of the future observations (see, {\it e.g.}, Refs.~\cite{pheno_tensor1,pheno_tensor2,
pheno_tensor3,pheno_tensor4}). In this work, we will focus on scalar modes, which are more relevant from 
the observational point of view but which are more demanding to deal with at the theoretical level because of subtle gauge-invariance issues and hypersurface deformation algebra closure conditions. \\
Two main types of  quantum corrections are expected at the effective level of LQC. The first one comes from 
the fact that loop quantization is based on holonomies, {\it i.e.} using exponentials of the connection rather than direct connection components. The second type of corrections arises for inverse powers of the densitized triad, which, when quantized, becomes an operator without zero eigenvalue in its discrete spectrum, thus avoiding the divergence.  As the status of ``inverse volume" corrections is not fully clear, due to the fiducial volume dependence, this work focuses on the holonomy term alone which has a major influence on the background equations and is better controlled \cite{eucl3}. In this framework, we will consider the Euclidean phase predicted by LQC \cite{eucl1, tom2}, and put, as advocated in Ref.~\cite{linda_pk}, initial conditions in the remote past of the contracting branch of the universe (this choice can be questioned and other proposals have been considered \cite{silent1,silent2}).\\
It is worth noticing, that an alternative attempt regarding the cosmological perturbations in LQC have recently 
been presented (see Refs.~\cite{abhay_new}). In this approach, quantum fields are considered on a
homogeneous quantum background, based on the methods developed in Ref.~\cite{asht_lew_2009}. Because the gauge-invariant variables for perturbations are fixed to be the classical ones, the Euclidean phase characterized by the elliptic nature of the equations of motion does not occur.  However, the consistency of the effective dynamics emerging from this formulation remains an open issue.\\

The key ingredient of this work is the existence of an Euclidean phase around the bounce. A completely rigorous study of the physical consequences would require a full understanding of quantum field theory (QFT) in curved Euclidean spaces and at the junction hypersurface with the Lorentzian manifold. This is obviously far beyond the scope of this article. Especially when taking into account that even on a well behaved Lorentzian dynamical space, QFT is not without ambiguities and many issues remain open. The central methodology of this study is to define physical quantities in the initial classical Lorentzian space (the contracting branch before the bounce), to evolve them in a mathematically rigorous way through the Euclidean zone, and to calculate observables in the second Lorentzian phase (the expanding branch we live in) where the physical meaning is clear again. Although, by definition, there is no time anymore in the Euclidean phase, one can still rely on general covariance and the corresponding structure has a well-defined canonical formulation using hypersurface deformations. We don't claim that this is the only way to address this situation. There are clearly other possible ways to deal with this speculative new phenomenon. However, it seems to be a quite natural and reasonable first assumption. In addition, this is the methodology that has been used up to now to evaluate the tensor spectrum in the deformed algebra approach to LQC. It is therefore important to also derive the scalar spectrum following the very same methodology, at least for a meaningful comparison.\\

In the following section, we first remind the basis of the deformed algebra approach used in this study. 
In section III, we summarize some important features of the background dynamics in LQC. The equation of 
motion for scalar perturbations is derived in section IV. In section V, different ways of choosing initial conditions 
for perturbations are presented. Section VI is devoted to the analysis of the scalar power spectrum. Concluding remarks are given in section VII.
\section{Deformed algebra}
In the canonical formulation of general relativity, the Hamiltonian is a sum of three constraints, 
\begin{equation}
H_{\rm G}[N^i,N^a,N] = \frac{1}{2\kappa } \int_{\Sigma}d^3x \left( N^i C_i+N^a C_a+N C\right) \approx 0,
\nonumber 
\end{equation}
where $\kappa = 8\pi G$, $(N^i,N^a,N)$ are Lagrange multipliers,  $C_i$ is 
 the Gauss constraint,  $C_a$ is the diffeomorphism  constraint and $C$ is 
the scalar constraint.  The equality denoted as "$\approx$" is to be understood as an equality on the surface of 
constraints ({\it i.e.} a weak equality). It is convenient to define the corresponding smeared 
constraints, 
\begin{eqnarray}
\mathcal{C}_1 &=& G[N^i] =\frac{1}{2\kappa } \int_{\Sigma}d^3x\ N^i C_i, \\
\mathcal{C}_2 &=& D[N^a] =\frac{1}{2\kappa } \int_{\Sigma}d^3x\ N^a C_a, \\
\mathcal{C}_3 &=& S[N] =\frac{1}{2\kappa } \int_{\Sigma}d^3x\ N C,
\end{eqnarray}
such that $H_{\rm G}[N^i,N^a,N] = G[N^i] +D[N^a]+S[N]$. The Hamiltonian is a 
total constraint which vanishes for all multiplier functions $(N^i,N^a,N)$. The time derivative
of the Hamiltonian constraint vanishes also weakly and therefore
the Hamilton equation, $\dot{f}=\{f,H_{\rm G }[M^i,M^a,M]\}$, leads to
\begin{equation}
\left\{H_{\rm G }[N^i,N^a,N], H_{\rm G }[M^i,M^a,M]\right\} \approx 0. \label{HH}
\end{equation}
As the Poisson brackets are linear, the condition 
(\ref{HH}) is satisfied if the smeared constraints belong to a first class algebra,
\begin{equation}
\{ \mathcal{C}_I, \mathcal{C}_J \} = {f^K}_{IJ}(A^j_b,E^a_i) \mathcal{C}_K, \label{algebra}
\end{equation}
where the ${f^K}_{IJ}( A^j_b,E^a_i)$ are structure functions which depend on the Ashtekar variables  $(A^j_b, E^a_i)$. The algebra closure is fulfilled at the classical level due to general covariance. The algebra must also be closed at the quantum level. Otherwise the system might escape from the surface of constraints, leading to an unphysical behavior.  In addition, as shown in Ref.~\cite{Nicolai:2005mc}, the algebra of effective quantum constraints should be strongly closed (that is, off shell closure must be considered). This means that the relation (\ref{algebra}) should hold in the whole kinematical phase space, and not only on the surface of constraints (corresponding to on shell closure). 
When the constraints are quantum-modified by the holonomy corrections, the resulting 
Poisson algebra might not be closed,  
\begin{equation}
\{ \mathcal{C}^Q_I, \mathcal{C}^Q_J \} = {f^K}_{IJ}(A^j_b,E^a_i) \mathcal{C}^Q_K+
\mathcal{A}_{IJ},
\end{equation}
where $\mathcal{A}_{IJ}$ stands for the anomaly term which can appear due to the 
quantum modifications and the superscript `$Q$' indicates that the constraints are quantum corrected. 
The consistency (closure of the algebra) requires that all $\mathcal{A}_{IJ}$ should vanish. 
Remarkably, the conditions, $\mathcal{A}_{IJ}=0$, lead to restrictions on  the form of the quantum 
corrections and determine them uniquely under natural assumptions. \\
The issue of anomaly freedom for the algebra of cosmological perturbations was extensively studied for  inverse-triad corrections. It was demonstrated that this requirement can be fulfilled at first order in perturbations for scalar \cite{Bojowald:2008gz,Bojowald:2008jv}, 
vector \cite{Bojowald:2007hv} and tensor perturbations \cite{Bojowald:2007cd}. Predictions for the power spectrum of cosmological perturbations were  performed \cite{Bojowald:2010me}, leading to constraints on some parameters of the model by the use of observations of the cosmic microwave background radiation (CMB) \cite{Bojowald:2011hd}. It was also considered for holonomy corrections and vector modes in Ref.~\cite{tom1} and for scalar modes in Ref.~\cite{tom2} (the full analysis with inverse-triad and holonomy terms was performed in Ref.~\cite{tom3}). It was shown in Ref.~\cite{eucl2} that there exists a single modification of the algebra structure that works for {\it all} kinds of modes, thus emphasizing the consistency of the theory. It is also important to underline that the matter content plays a role in removing degeneracies.
Even if the calculations carried out in the above-mentioned articles are quite laborious, the guiding idea behind is very simple. Each time a  $\bar{k}$ factor, defined as the mean value of the Ashtekar connection $A^i_a$, appears, it is replaced by 
\begin{equation}
\bar{k} \rightarrow \frac{\sin(n\bar{\mu} \gamma \bar{k})}{n\bar{\mu}\gamma},
\label{kbar}
\end{equation}
where $n$ is some unknown integer and $\bar{\mu}$ is the coordinate size of a loop. The full perturbations 
have to be calculated up to the desired order, the Poisson brackets are then explicitly calculated and the 
anomalies are cancelled by counter-terms required to vanish in the classical limit. The neat result is that the 
algebra of effective constraints is deformed with respect to its classical counterpart. It takes the following form:  
\begin{eqnarray}
\left\{D[M^a],D [N^a]\right\} &=& D[M^b\partial_b N^a-N^b\partial_b M^a], \nonumber \\
\left\{D[M^a],S^Q[N]\right\} &=& S^Q[M^a\partial_b N-N\partial_a M^a],  \nonumber  \\
\left\{S^Q[M],S^Q[N]\right\} &=& \Omega
D\left[q^{ab}(M\partial_bN-N\partial_bM)\right], \nonumber
\end{eqnarray}
where $\Omega$ is the deformation factor that plays a crucial role in the following. 
It is given by $\Omega=1-2\rho/\qsubrm{\rho}{c}$ where $\rho$ is the density of the 
Universe and $\qsubrm{\rho}{c}$ is the critical density expected to be close to the Planck density. 
In Lorentzian General Relativity $\Omega=1$. When $\Omega<0$ the structure of
 space-time becomes Euclidean. (Strictly speaking space-time is Lorentzian or Euclidean only 
if $\Omega = \pm1$ but the most important properties regarding physical consequences,
namely the existence of a causal structure and the general behavior of the solutions for wave equations, 
only depend on the sign of $\Omega$ and not on its precise value \cite{silent2}. It therefore makes sense to speak of Lorentzian or Euclidean phases.) \\
Interestingly, this conclusion has strong links with results often postulated (for technical reasons, 
notably a better behavior of path integrals) in quantum cosmology, but it appears here as a real dynamical prediction of the theory. In standard quantum cosmology one usually deals with an amplitude
\begin{equation}
<\phi_2,t_2|\phi_1,t_1>=\int d[\phi]e^{I[\phi]},
\label{eq:qc}
\end{equation}
where $I[\phi]$ is the action of the field configuration $\phi(x,t)$, and $d[\phi]$ is a measure on the space of field configurations. The integrand of Eq.\eqref{eq:qc} has a rapidly oscillating phase, and the path integral, in general, does not converge. This is why the time is rotated clockwise by $\pi / 2$ so that $I[\phi] \rightarrow \tilde{I}[\phi]\equiv-iI[\phi]$. The integrand in the resulting Euclidean path integral is now exponentially damped, and the integral generically converges. Then, one can analytically continue the amplitude in the complex $t$-plane back to real values. Importantly, a quantum field theory machinery has been developed in this framework and the interested reader can, {\it e.g.}, consider Ref.~\cite{fang}.
Of course, there are also links with  the 
Hartle-Hawking proposal \cite{hartle}. But the Euclidean phase appears in the model considered in this article in a fundamentally dynamical way since the Poisson bracket between Hamiltonian constraints varies continuously from a positive to a negative expression. The ``spirit" of the Hartle-Hawking proposal, translated in the framework considered here, has been studied in \cite{silent1}. Importantly, the appearance of an Euclidean phase was 
also independently derived from another approach to LQC in Ref.~\cite{ed}. Many quantum gravity 
approaches seem to predict the existence of a silent surface ($\Omega=0$) where light 
cones are completely squeezed, on each ``side" of the Euclidean phase. This is also a clear realization 
of the BKL conjecture (see, {\it e.g.} Ref.~\cite{jakub_silence}). Arguments are given in Ref.~\cite{eucl3} showing 
that the change from a hyperbolic to an elliptic type of equations in LQC should be understood as a 
true change of signature (that has been missed before because homogeneous models cannot probe it) 
and not just a tachyonic instability.
It should also be emphasized that the deformed algebra approach is grounded in avoiding 
gauge issues. Many approaches make a gauge fixing. In most cases, gauge fixing before quantization 
is known to be harmless, but the situation is different in general relativity. The constraints we are 
considering are much more complicated functions than, for example, the Gauss constraint of Yang--Mills 
theories: it is therefore likely that the constraints receive significant quantum corrections. If the constraints 
are quantum corrected, the gauge transformations they generate are, as we have shown, not of the 
classical form. Gauge fixing before quantization might then be inconsistent because one would fix the 
gauge according to transformations which subsequently will be modified. In addition, in the present case, 
the dynamics is part of the gauge system. A consistent theory must therefore quantize gauge transformations 
and the dynamics at once. It is not correct to fix one part (the gauge) in order to derive the second part 
(the dynamics) in an unrestricted way.  The subtle consistency conditions associated with the covariance 
of general relativity are encoded in the first class nature of its system of constraints. Here, great care is taken 
in not breaking this consistency.\\
The  equations of motion derived in this framework are still covariant under the deformed algebra replacing 
classical coordinate transformations. The corresponding quantum space-time structure is obviously not Riemannian (there is no line element in the usual sense), but has a well-defined canonical formulation using hypersurface deformations.
\section{Background evolution}
\begin{figure}[H]
\includegraphics[scale=0.41]{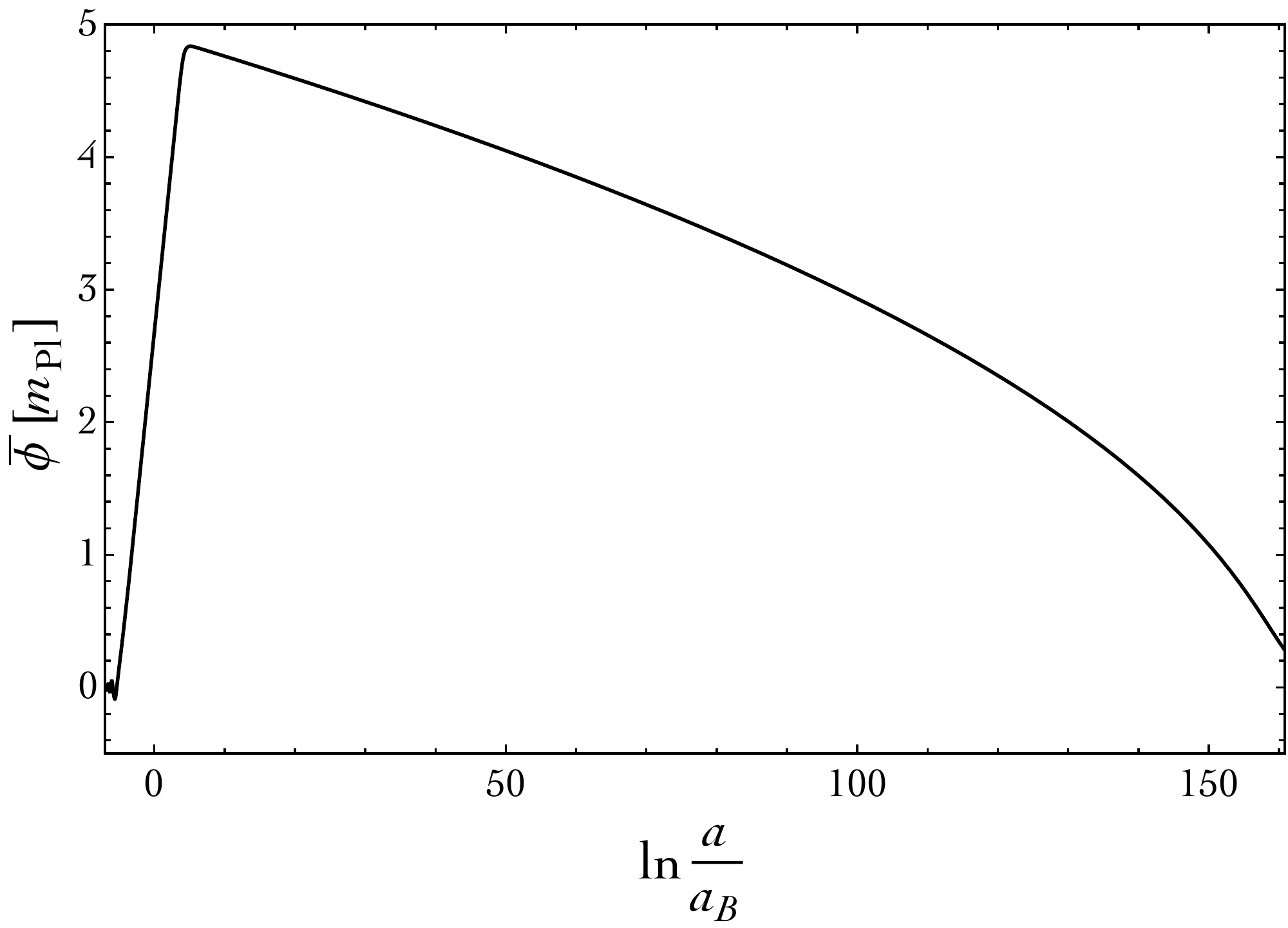}
\caption{Evolution of the scalar field as a function of the number of $e$-folds $N:=\pm \ln a/\qsubrm{a}{B}$, 
with $m = 1.2 \times 10^{-6} \qsubrm{m}{Pl}$. The zero on the horizontal axis corresponds to the
bounce when $\qsubrm{a}{B} = 1$. This solution is such that $\qsubrm{x}{B} = 3.55 \times 10^{-6}$  
(obtained with $\alpha = 17 \pi/4 +1$ and $ \qsubrm{\theta}{0} = 5.11$). The evolution 
is stopped at the end of inflation when $\phi = 1/ \sqrt{4 \pi} ~ \qsubrm{m}{Pl}$.}
\label{fig:bkg}
\end{figure}

At the background level, the change of signature cannot be probed/detected. This is obvious for two reasons. First, because the relative sign between temporal and spatial derivatives cannot be identified. Second, because the Poisson bracket between Hamiltonian constraints then trivially vanishes.
The evolution of the cosmological background is studied at the effective level with holonomy corrections. 
The background geometry is described by the homogeneous, isotropic and flat configuration 
parametrized by the scale factor $a$. The dynamics of the background is governed by the quantum-corrected 
Friedmann equation
\begin{equation}
H^2=\frac\kappa3\rho\left(1-\frac{\rho}{\qsubrm{\rho}{c}}\right),
\label{eq:mfe}
\end{equation} 
derived in Ref.~\cite{ImprovedDynamics}, where $H = \dot{a}/a$ is the Hubble rate 
in cosmic time, $\rho$ is the energy density of the content of the universe and 
$\qsubrm{\rho}{c}$ denotes its maximal value attained at the bounce. The dot denotes 
a derivative w.r.t. cosmic time. Obviously, the physical interpretation of the background equations makes sense only in the Lorentzian phase, that is when $\rho<\rho_c/2$, but technically one can still determine the manifold structure for $\rho_c/2<\rho<\rho_c$, that is in the Euclidean phase. An alternative view of our approach which solves all interpretation difficulties and which leads to the very same result, is to consider that there is no {\it real} change of signature: the background evolves in a standard way (although, of course, according to the modified Friedmann equation) and the change of sign of the $\Omega$ factor entering the propagation equation of perturbations is just a tachyonic instability, as well known to appear, for example, in Gauss-Bonnet gravity when the curvature invariant is non-minimally coupled with a scalar field. We prefer not to favor this view because, as detailed in \cite{eucl3}, there are hints that the phenomenon is deeper but one is free to see things in this way. Planck units are used throughout this article with 
$\qsubrm{m}{Pl} = 1/ \sqrt{G} \approx 1.22 \cdot 10^{19}$ GeV.
Furthermore, we consider a single massive scalar field, $\phi$, with a quadratic potential 
$V = m^2 \phi^2 /2$, as the matter content of the Universe. This choice is made for simplicity. It allows easy comparisons  with other works and  generates a phase 
of slow-roll inflation. Even if this potential is not favored by  current observational data  
\cite{Planck2015}, it still serves as a valuable toy model for studying the phase of 
inflation in different frameworks. Moreover, taking into account more subtle effects, 
{\emph e.g.} the quantum gravitational corrections considered here, might improve the
status of the quadratic potential in the light of the observational data. \\
Splitting the field $\phi = \bar{\phi} + \delta \phi$ into a background part, 
$\bar{\phi}$, and a perturbed part, $\delta \phi$, the Klein-Gordon equation for 
the background reads
\begin{equation}
\ddot{\bar{\phi}} +3H\dot{\bar{\phi}}+m^{2}\bar{\phi}=0.\label{eq:kge}
\end{equation}
A first analysis of this model has already been studied in Ref.~\cite{Vandersloot}; 
a detailed analysis of the background equations can be found in Ref.~\cite{boris}. 
Here, we only summarize the main features of the background dynamics. 
The field evolution can be characterized by two dynamical parameters \cite{pheno_tensor3}, 
the potential energy parameter, $x$, and 
the kinetic energy parameter, $y$, defined as
\begin{equation}
x := \frac{m \bar{\phi}}{\sqrt{2 \qsubrm{\rho}{c}}}, ~~~~~y := \frac{\dot{\bar{\phi}}}{\sqrt{2 \qsubrm{\rho}{c}}}.
\end{equation}
Then the total energy density can be written as  $\rho = \qsubrm{\rho}{c} (x^2 + y^2)$. Eqs. \eqref{eq:mfe} and 
\eqref{eq:kge} can then be recast as
\begin{eqnarray}
\begin{cases}
\dot{H} & =- \kappa \qsubrm{\rho}{c}y^{2}\left(1-2x^{2}-2y^{2}\right),\\
\dot{x} & =my,\\
\dot{y} & =-3Hy-mx,
\end{cases}\label{eq:sys}\end{eqnarray}
showing that there are two timescales involved in this system: one is given by $1/m$ and corresponds 
to the classical evolution of the field, the other one is $1/\sqrt{3 \kappa \qsubrm{\rho}{c}}$ and 
corresponds to the quantum regime of the evolution. The ratio of these two timescales is
\be
\Gamma := \frac{m}{\sqrt{3 \kappa \qsubrm{\rho}{c}}}.\label{eq:c1}
\end{equation}
 According to standard assumptions of slow-roll inflation with a quadratic 
potential, the value of the mass $m \simeq 1.2 \times 10^{-6} \qsubrm{m}{Pl}$ is preferred 
in the light of the observational data from the Planck satellite (see Ref.~\cite{Planck2015}). 
The critical energy density at the bounce is given by $\qsubrm{\rho}{c} = 0.41\ \qsubrm{m}{Pl}^4$, 
which is exactly the upper bound of the spectrum of the energy density operator \cite{Ashtekar_Review}. 
These values lead to $\Gamma \simeq 2 \cdot 10^{-7}$. Hence, one can safely assume that $\Gamma \ll 1$, 
ensuring that  the evolution splits into three phases: (i) a classical pre-bounce contracting phase, 
(ii) the bouncing phase and (iii) a classical expanding phase after the bounce (slow-roll inflation), 
see Ref. \cite{boris} for details. Initial conditions, $\left\lbrace \qsubrm{a}{0}, \qsubrm{x}{0}, 
\qsubrm{y}{0} \right\rbrace$, are set in the remote past of the contracting phase when the energy 
density is very small compared to the critical energy density, {\it i.e.}
\begin{equation}
\sqrt{\frac{\qsubrm{\rho}{0}}{\qsubrm{\rho}{c}}} \ll \Gamma.
\label{initialenergydensity}
\end{equation}
It is convenient to use polar coordinates for the potential and kinetic energy parameters
\begin{align}
x(t) &= \sqrt{\frac{\rho}{\qsubrm{\rho}{c}}} \sin (m t + \qsubrm{\theta}{0}), \\
y(t) &= \sqrt{\frac{\rho}{\qsubrm{\rho}{c}}} \cos (m t + \qsubrm{\theta}{0}),
\label{xyoscillation}
\end{align}
where $\qsubrm{\theta}{0}$  is the initial phase between the share of potential energy and kinetic energy. 
In order to select different background evolutions independently of the small oscillatory behavior of the 
solutions, the following parametrization shall be used:
\begin{equation}
\sqrt{\frac{\qsubrm{\rho}{0}}{\qsubrm{\rho}{c}}} = \frac{\Gamma}{\alpha} \left( 1 
- \frac{\sin(2 \qsubrm{\theta}{0})}{4 \alpha} \right)^{-1},
\end{equation}
where $\alpha$ is a number large enough such that \eqref{initialenergydensity} holds. To each 
phase, $\qsubrm{\theta}{0}$, corresponds a specific value of the potential energy parameter at the 
bounce $\qsubrm{x}{B} $. As shown in  Ref.~\cite{DurationInflationLinda}, for a mass $m =1.21 \times 10^{-6} \qsubrm{m}{Pl}$, 
the favored value for $\qsubrm{x}{B}$ is $ 3.55 \times 10^{-6}$. This solution for the background 
dynamics features only a tiny amount of deflation before the bounce as shown in Fig.~\ref{fig:bkg}. 
In general, we will chose the normalization of the scale factor at the bounce as $\qsubrm{a}{B} = 1$. 
The plots and spectra are presented as functions of the number of $e$-folds 
$N:= \pm \ln (a/\qsubrm{a}{B})$, that have to elapse until the bounce (negatively valued) and 
that have elapsed after the bounce (positively valued) respectively.
\section{Equation of motion for scalar modes}
The equation of motion for  scalar modes in the deformed algebra approach is derived from the particular form of the Hamiltonian constraint. In Ref.~\cite{eucl2}, the gravitational part of the Hamiltonian constraint has been analyzed and reads (up to quadratic order) 
\be 
H[N] = \int_\Sigma d^3x \left[ \barn\,( \mathcal{H}^{(0)}+ \mathcal{H}^{(2)}) + 
\delta N\, \mathcal{H}^{(1)}\right],
\ee
where
\begin{eqnarray}
2 \kappa \; \mathcal{H}^{(0)} &=&- 6 \sqrt{\barp} \bark^2 ~, \\
2 \kappa\; \mathcal{H}^{(1)}& =&- 4 \sqrt{\barp} \delta K^d_d -
\frac{\bark^2}{\sqrt{\barp}} \delta E^d_d + \frac{2}{\sqrt{\barp}}  \partial^j \partial_c \delta E^c_j,\\
2 \kappa\;  \mathcal{H}^{(2)} &=& - 2 \frac{\bark}{\sqrt{\barp}} \delta K^i_a\delta E^a_i \nonumber \\ 
\hspace{-1cm} &+&  \sqrt{\barp}\,  \big( \delta^b_i  \delta K^i_a \delta^a_j \delta K^j_b - 
\delta^a_i \delta K^i_a \delta^b_j \delta K^j_b \big)\, \nonumber \\
\hspace{-1cm} &+&  \frac{1}{4} \frac{\bark^2}{\barp^\frac{3}{2}}
\big( \delta^i_a \delta E^a_i \delta ^j_b \delta E^b_j 
- 2 \delta ^j_a \delta E^a_i \delta ^i_b \delta E^b_j \big) \nonumber \\
\hspace{-1cm} &+&  \frac{1}{\barp^{\frac{3}{2}}} Y^{kjil}_{bdc} ~\epsilon^{ab}_k \; 
\partial_a \left( \delta E^d_j \partial_i \delta E^c_l\right) \nonumber \\
\hspace{-1cm} &+& \frac{1}{\barp^\frac{3}{2}} Z^{cidj}_{ab} \; 
\big( \partial_c \delta E^a_i\big)\big(\partial_d \delta E^b_j\big). \label{BIGMAMMA}
\end{eqnarray}
Here, $\bar{p}$ is the mean value of the densitized triad $E^a_i$ and $\bar{k}$ was 
defined in Eq.~\eqref{kbar}. The term $Z^{cidj}_{ab}$ depends on the kind of modes 
considered (scalar, vector or tensor):
\be
Z^{cidj}_{ab} = \begin{cases} 
 \delta_{ab} \delta^{ij} \delta^{cd}&\mbox{for tensor modes,}  \\
~~~~~ 0 &\mbox{for vector modes,}  \\
- \frac{1}{2} \delta^c_a \delta^d_b \delta^{ij} &\mbox{for vector modes.}
\end{cases}
\ee
Lastly, $Y^{kjil}_{bdc}$ is a complicated expression whose form is not relevant here.  
Based on this, the holonomy quantum corrections can be accounted for and the 
Mukhanov-Sasaki equation of motion for gauge-invariant perturbations can be 
calculated \cite{tom2}. In conformal time it is given by
\begin{equation}
\qsubrm{v}{S}'' - \Omega \, \nabla ^2 \qsubrm{v}{S} - \frac{\qsubrm{z}{S}''}{\qsubrm{z}{S}} \qsubrm{v}{S} = 0,
\label{EoMDeformedAlgebra}
\end{equation}
with
\be
\qsubrm{v}{S}:=\sqrt{\barp}\,\left(\delta\phi + \frac{{\bar\phi}'}{\Hc}\Phi\right)
\hskip4mm \mbox{and}\hskip4mm
 \qsubrm{z}{S} \;:= \; \sqrt{\barp}\, \frac{{\bar{\phi}}'}{\Hc}.
\end{equation}
The variable $\Phi$ denotes the gauge invariant Bardeen potential taking into account the metric 
perturbations, whereas $\phi$ represents the massive scalar field. $\Hc$ is the conformal Hubble 
parameter. The Mukhanov equation of motion  \eqref{EoMDeformedAlgebra} reduces to the classical 
equation when $\Omega \rightarrow 1$. Note that for FLRW cosmologies, in conformal time, 
$\sqrt{\bar{p}} = a$. On the quantum-modified background discussed in the previous section, we can 
evaluate the evolution of the Mukhanov variable $\qsubrm{v}{S}$. For simplicity we will omit the index 
`${\rm S}$' in the following, assuming that it is clear that $v$ denotes the scalar perturbation variable.\\ 
Using the Fourier space decomposition of the $v({\bf x},\eta)$ field,
\begin{equation}
v({\bf x},\eta) = \int \frac{d^3k}{(2\pi)^{3/2}} v_{\bf k }(\eta) e^{i {\bf k \cdot x}},
\end{equation}
one gets a set of ordinary differential equations for the Fourier components $v_{\bf k }$.
Due to the isotropy of space, the ${\bf k }$-vector of $v_{\bf k }$ might be simplified to 
the absolute value dependence $v_k$, where $k := \sqrt{\bf k \cdot k}$. The  $v_k$ function 
is called a mode function.  
Instead of using the conformal time dependence, it is often (due to technical reasons) 
convenient to switch to cosmic time $t$ in the numerical computations. With $t =  \int a \cdot d\eta$, 
the Mukhanov equation of motion reads
\begin{equation}
\ddot{v}_k + H \dot{v}_k + f_{k}^{\scriptscriptstyle{(v)}}(t) v_k = 0,
\label{MukhFourier}
\end{equation}
with  $z= a \frac{\dot{\bar{\phi}}}{H}$ and 
\be
f_{k}^{\scriptscriptstyle{(v)}}(t)  :=  \Omega\frac{k^2}{a^2} - \frac{\dot{z}}{z} H - \frac{\ddot{z}}{z},
\label{vEffpot}
\ee
being the effective frequency term.
In order to derive the primordial power spectrum after inflation one would have to solve Eq.~\eqref{MukhFourier} for every 
mode $k$ for all times from $\qsubrm{t}{init}$ until $\qsubrm{t}{end}$ where $\qsubrm{t}{init}$ is the initial 
starting point, set in the remote past as we will see later, and $\qsubrm{t}{end}$ denotes the time at the 
end of the inflationary phase.  This requires a numerical integration. However, the Mukhanov variable $v$,
 cannot be used for the whole integration because of a non-physical singularity occurring at the bounce. 
Let us describe how to bypass this difficulty by using the change of variable.  
We introduce $h_k := v_k/a$ for every $k$, so that \eqref{MukhFourier} becomes
\begin{equation}
\ddot{h}_k + 3H \dot{h}_k + f_{k}^{\scriptscriptstyle{(h)}}(t) h_k = 0
\label{hFourier}
\end{equation}
with 
\be
f_{k}^{\scriptscriptstyle{(h)}}(t) :=  \Omega\frac{k^2}{a^2}  + m^2 + m^2 \kappa  ~ 
\Omega \frac{\dot{\bar{\phi}} \bar{\phi}}{H} - 2 \left( \frac{\dot{H}}{H} \right)^2 + \frac{\ddot{H}}{H} .
\label{hEffPot}
\ee
In the numerical computation this second order differential equation is replaced by the following first order system: 
\begin{eqnarray}
\begin{cases}
\dot{h}_k &= (1/a) g_k,\\
\dot{g}_k & = -2 H g_k + f_{k}^{\scriptscriptstyle{(h)}}(t) a~ h_k.
\end{cases}\label{heqsyst}\end{eqnarray}
The numerical integration of \eqref{heqsyst} is performed for 
$t \in [\qsubrm{t}{init},t_{\scriptscriptstyle{h\rightarrow \mathcal{R}}}]$ 
where $t_{\scriptscriptstyle{h\rightarrow \mathcal{R}}}$ is before the bounce. Since the effective frequency terms, \eqref{hEffPot} or \eqref{vEffpot}, depend on inverse powers of $H$, the differential equations have a generic singularity at 
the bounce, when the Hubble parameter vanishes. Nonetheless, this singularity is not physical, which can be seen by analyzing the physical scalar curvature,
\be
\mathcal{R} := \frac{v}{z}. 
\ee
Using this variable, one can rewrite Mukhanov's equation of motion in Fourier space as
\be
\ddot{\mathcal{R}}_k - \left( 3H + 2 m^2 \frac{\bar{\phi}}{\dot{\bar{\phi}}} + 
2 \frac{\dot{H}}{H} \right) \dot{\mathcal{R}}_k + \Omega\frac{k^2}{a^2} \mathcal{R}_k = 0.
\label{eomR}
\ee
When approaching the bounce, $H$ tends to zero and $\Omega$ to minus one. It should be noticed that this equation is mathematically well behaved, even when $\Omega <0$, that is in the Euclidean-like phase. Obviously, in the Euclidean phase, $k$ loses its usual interpretation and the Fourier transformation has no intuitive physical sense. This is, however, already true on a Lorentzian curved manifold: the space/ time splitting -- or positive/ negative frequency splitting -- is, in general, ill defined: one usually relies on static boundaries that allow a physical choice. The ``bet" of this study is fundamentally the same: quantities are defined in the remote past  of the contracting branch when the signature is the usual one, and when quantum effects are negligible. Then, they are mathematically propagated through the bounce where they lose a clear meaning. Finally, the observables are computed in the expanding Lorentzian phase when physics is again under control. This is a questionable approach, but, in our opinion, a reasonable one at this stage.

During the bouncing phase, $\dot{\bar{\phi}} \simeq \sqrt{2 \qsubrm{\rho}{c}} \gg 
m \bar{\phi}$, so that the equation of motion reduces to
\begin{equation}
\ddot{\mathcal{R}}_k - 2 \frac{\dot{H}}{H} \dot{\mathcal{R}}_k - \frac{k^2}{a^2} \mathcal{R}_k \simeq 0.
\label{eq:EoMBounce}
\end{equation}
According to the analytical expression for $\bar{\phi}$ and $\dot{\bar{\phi}}$ around the bounce developed 
in Ref. \cite{boris}, this gives 
\begin{equation}
\ddot{\mathcal{R}}_k - \frac{2}{t- \qsubrm{t}{B}} \dot{\mathcal{R}}_k - k^2 \mathcal{R}_k \simeq 0,
\end{equation}
when we restrict ourselves to the first order in $(t- \qsubrm{t}{B})$. The space of solutions to 
this differential equation is spanned by the two independent functions,
\begin{align*}
{\mathcal{R}}^{\scriptscriptstyle{(1)}}_{k} &= 
\left[ \sinh \left( k (t - \qsubrm{t}{B})\right) - k (t - \qsubrm{t}{B}) \cosh\left(k (t - \qsubrm{t}{B}) \right) \right], \\
\mathcal{R}^{\scriptscriptstyle{(2)}}_{k} &= 
\left[  \cosh \left( k (t - \qsubrm{t}{B})\right) - k (t - \qsubrm{t}{B}) \sinh \left(k (t - \qsubrm{t}{B}) \right)  \right].
\end{align*}
These solutions show  an obviously regular behavior at the bounce. But it should be noticed that \eqref{eomR} runs into trouble away from the bounce due to the time 
derivative of the potential $\dot{\bar{\phi}}$ appearing in the denominator of the friction term. During the classical contracting and expanding phases,  $\dot{\bar{\phi}}$ oscillates around a null value, causing the break-down of the numerical integration of the differential equation 
\eqref{eomR} of $\mathcal{R}$. For this reason one has to switch twice between Eq.~\eqref{heqsyst} and Eq.~\eqref{eomR} during the numerical computations. 
For $t \in [\qsubrm{t}{init},t_{\scriptscriptstyle{h\rightarrow \mathcal{R}}}]$ and $t \in [t_{\scriptscriptstyle{\mathcal{R} \rightarrow h}}, \qsubrm{t}{end}]$, where 
$t_{\scriptscriptstyle{h\rightarrow \mathcal{R}}} < \qsubrm{t}{B} < t_{\scriptscriptstyle{\mathcal{R} \rightarrow h}}$, the differential equation for $h$, namely \eqref{heqsyst}, must be used. Whereas for 
$t \in [t_{\scriptscriptstyle{h\rightarrow \mathcal{R}}}, t_{\scriptscriptstyle{\mathcal{R} \rightarrow h}}]$, 
it is the equation for $\mathcal{R}$, Eq. \eqref{eomR}, that has to be integrated.  The exact choice 
of the transition points is irrelevant as long as they do not approach one of the singularity points. 
\section{Initial conditions}
\begin{figure}[H]
\includegraphics[scale=0.42]{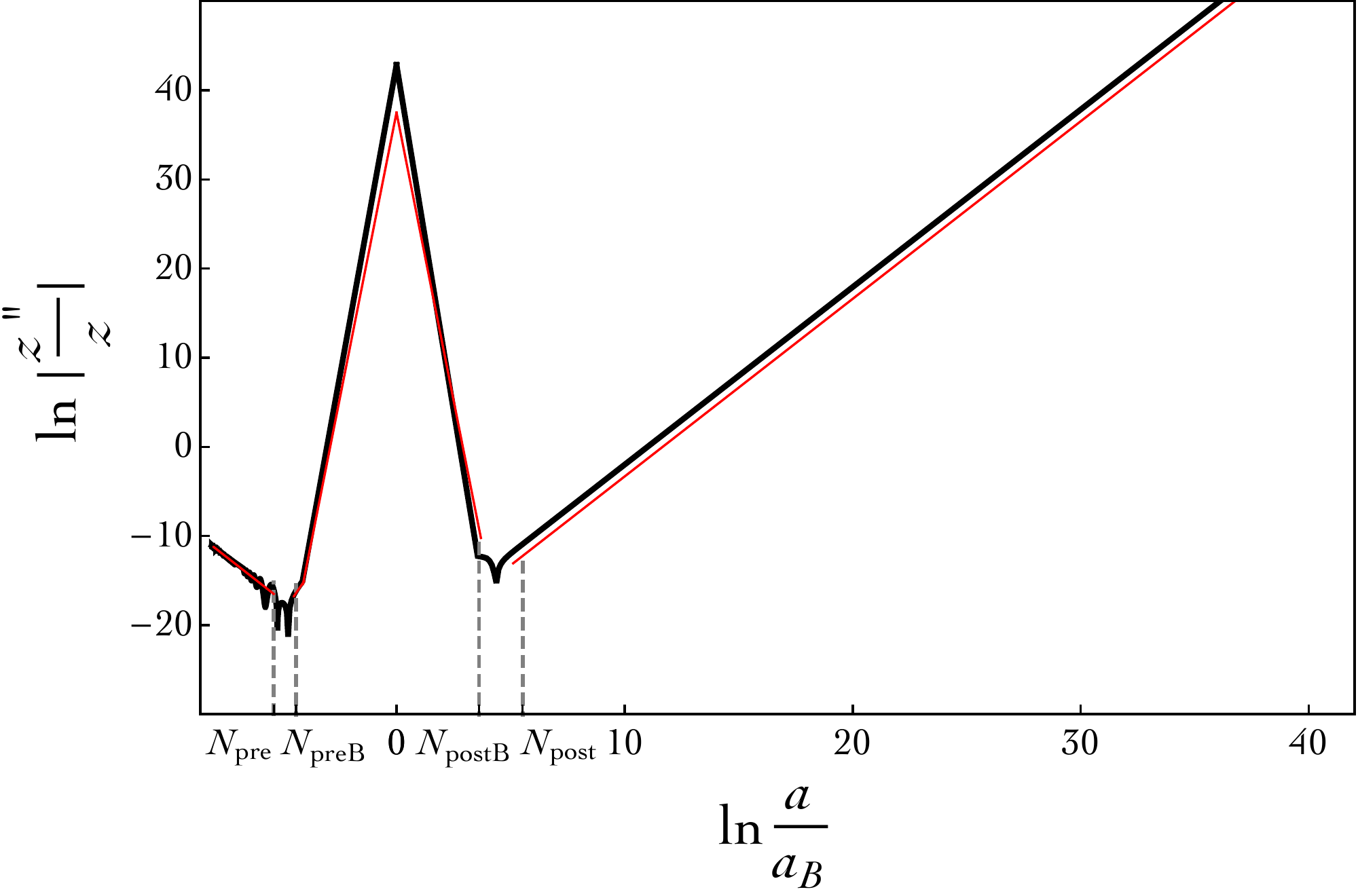}
\caption{Evolution of the cosmological term $z''/z$ as a function of the number of $e$-folds 
$\ln(a/\qsubrm{a}{B})$, with $m = 1.2 \times 10^{-6} \qsubrm{m}{Pl}$. The parameters for the 
background are set as in Fig.~\ref{fig:bkg}. During the pre-bounce contracting phase and slow-roll 
inflation $\ln(z''/z)\propto \pm 2\ln (a/\qsubrm{a}{B}) + const.$}
\label{fig:effpot}
\end{figure}

The considered equations of motion for the scalar perturbations (Eq. \eqref{EoMDeformedAlgebra} 
or Eq. \eqref{MukhFourier}) might be considered -- at the effective level -- as the quantum ones. This is due to the presence of the factor $\Omega$, being a result of the quantum gravitational effects. The quantum effects taken into account here are, however, only those which modify the background degrees of freedom. The inhomogeneous degrees might be (and are), treated classically in the perturbative regime under consideration. In order to see it, let us consider the extrinsic curvature $K^i_a$ which is exponentiated to the form of a holonomy operator in the quantum theory. In the perturbative treatment we have $K^i_a=\bar{k}\delta^i_a+\delta K^i_a$, together with the condition $|\delta K^i_a|/\bar{k} \ll 1$. Path integration of $K^i_a$ leads to a factor of the form  $\gamma \bar{\mu} \bar{k}$ for the homogeneous contribution, which is of the order of unity in vicinity of the bounce. Full exponentiation of the background contribution to $K^i_a$ must be, therefore, kept over the evolution through the bounce. However, this is not necessary for sufficiently small perturbations, for which the condition $\gamma \bar{\mu} |\delta K^i_a| \ll 1$ might be satisfied even at the bounce. This allows for the expansion of the holonomy up to a linear contribution in $\delta K^i_a$ and treating the perturbative degrees of freedom in a classical manner. \\
As the phase space of the perturbative degrees of freedom is approximated by  the classical one, 
the canonical quantization procedure for the modes might be applied. The canonical quantization is an 
approximation, which is valid only for sufficiently small amplitudes of $\delta K^i_a$ and sufficiently 
large wavelengths (roughly greater than the Planck length) in the mode expansion. If the conditions 
are satisfied, the Fourier mode $v_{\bf k}(\eta)$ can be promoted to be an operator, such that 
in the Heisenberg picture
\begin{equation}
\hat{v}_{\bf k}(\eta) = v_k(\eta) \hat{a}_{\bf k} +v_k^*(\eta) \hat{a}^{\dagger}_{\bf-k}, 
\end{equation}
where $v_k(\eta)$ are the so-called mode functions satisfying the classical equation (\ref{MukhFourier}). 
The $\hat{a}^{\dagger}_{\bf k}$ and $\hat{a}_{\bf k}$ are the creation and annihilation
operators respectively, satisfying $[\hat{a}_{\bf k},\hat{a}^{\dagger}_{\bf q}] =\delta^{(3)}({\bf k-q})$. Using this commutation relation, one may show that the following condition  
\begin{equation}
v_k \frac{d v^{*}_k}{d\eta}-v^{*}_k \frac{d v_k}{d\eta} = i,
\label{Wronskian}
\end{equation}
called Wronskian condition, has to be satisfied in order to preserve the standard canonical structure. \\
Based on the above, the two-point correlation function for the scalar curvature field $\hat{\mathcal{R}}({\bold x},\eta)$ in the vacuum state $| 0 \rangle $ is given by 
\begin{eqnarray}
\langle 0|  \hat{\mathcal{R}}({\bold x},\eta) \hat{\mathcal{R}}({\bold y},\eta)| 0 \rangle 
=\int_0^{\infty} \frac{dk}{k} \mathcal{P}_{S}(k,\eta) \frac{\sin kr}{kr},   
\end{eqnarray}
where $r= |{\bold x}-{\bold y}|$ and the scalar power spectrum reads
\begin{equation}
\mathcal{P}_{S}(k,\eta) :=  \frac{k^3}{2 \pi^2} \frac{|v_k(\eta)|^2}{z^2}. 
\label{PowerSpectrum}
\end{equation}
The power spectrum carries all statistical information about the Gaussian scalar curvature field under 
consideration (non-linear effects are neglected in our analysis) and its determination for the 
LQC model discussed in the previous sections is a main goal of this study.\\
The equations of motion for the scalar mode functions $v_k(\eta)$ can be solved numerically, following 
the procedure presented in the previous section. For this purpose initial conditions for perturbations 
have to be set for every wavenumber $k$. In standard cosmology, it is common to set Cauchy initial 
conditions at some moment in time after the big bang singularity. In the present work, we set initial 
conditions  in the  pre-bounce phase. This is the natural choice if the bounce is a  phenomenon to be 
really understood as resulting from a causal evolution of the Universe, with time flowing in a unique 
direction. In addition, in the remote past of the contracting branch, the Universe is classical and quantum effects do not play any important role. This seems both technically more convenient (since the quantum dominated region still represents a quite unknown field of physics) and physically better motivated. In particular, as discussed in the previous sections, it has been shown in  Ref.~\cite{tom2} that the geometry of the universe in its quantum stage 
($\rho > \qsubrm{\rho}{c} /2$) might become Euclidean instead of being Lorentzian (the very notion 
of time obviously looses here its meaning). The physical consequences are still not perfectly well 
understood and setting initial conditions in the Euclidean phase would be the worst possible choice: we therefore focus on the classical contracting phase. Note that  another proposal, studied in Refs.~\cite{silent1,silent2}, is to set initial conditions at the surface of silence (or in a ``hybrid" way as advocated by a careful study of the Tricomi problem). We do not consider this hypothesis here. \\
The most simple and natural way to set initial conditions for a quantum oscillator is provided by the 
vacuum state $\ket{0}_k$ for every mode $k$ at some given moment in time. This sets a clear Cauchy 
initial value problem. However, it is well known that the notion of vacuum in an arbitrary curved spacetime 
is ambiguous since the definition of the usual ``instantaneous vacuum" is based on plane waves 
satisfying the differential equation of an harmonic oscillator with wavenumber $k$. In the case 
of scalar perturbations, as the effective frequency term depends non-trivially on time, it is more appropriate 
to use the  Wentzel-Kramers-Brillouin (WKB) approximation and the so-called ``adiabatic vacuum". 
In contrast with the ordinary instantaneous vacuum state, well known from quantum field theory, the 
adiabatic vacuum does not have to satisfy the differential equation of an harmonic oscillator in some
limit. The instantaneous-vacuum state is recovered as the first term of the WKB expansion. \\
For scalar perturbations in LQC the equation of motion is given by Eq.~\eqref{EoMDeformedAlgebra}.
In conformal time, this equation resembles the differential equation of an oscillator, 
\begin{equation}
v''_k + \qsubrm{k}{eff}^2(\eta) \qsubrm{v}{k} = 0,
\label{osciequation}
\end{equation}
with a time-dependent  wave number 
\begin{equation}
\qsubrm{k}{eff} (\eta) := \sqrt{\Omega(\eta) k^2 - \frac{z''}{z}(\eta)}.
\label{effpotdef}
\end{equation}
Recall that we set initial conditions in the remote past where $\Omega \sim 1$ is almost constant. 
Thus the $\Omega$-factor actually plays no role when addressing the issue of initial conditions.
The main idea of the WKB approximation is to use the following generic ansatz for the solutions to Eq.~\eqref{osciequation}:
\begin{eqnarray}
v_k(\eta) = c_1\cdot e^{i (\qsubrm{k}{eff} T) \cdot W_k(\eta)} 
               + c_2 \cdot e^{- i (\qsubrm{k}{eff} T) \cdot W_k(\eta)},
\end{eqnarray}
where the values of constants $c_1$ and $c_2$ are constrained according to the Wronskian 
condition (\ref{Wronskian}).  In the WKB approximation the functions $W_k(\eta)$ are expanded in terms of some small parameter $(\qsubrm{k}{eff} T)^{-1}$ where $T$ is the minimal time interval for which $\qsubrm{k}{eff}$, and its time derivatives, start to change substantially $(T \gg 1/\qsubrm{k}{eff})$. Then the WKB expansion reads
\begin{equation}
W_k(\eta) = \sum_{n=0}^{\infty} \left( \frac{i}{\qsubrm{k}{eff} T} \right)^n W_{k,n}(\eta).
\end{equation}
Introducing this ansatz into Eq.~\eqref{osciequation}, one gets the explicit expressions for the different orders, 
$n$, of $W_{k,n}$. The WKB approximation consists  in truncating the series after the first order, leading 
to the approximated solution for the mode functions
\begin{eqnarray}
v_k(\eta) &=& \frac{c_1}{\sqrt{\qsubrm{k}{eff}(\eta)}} e^{i \int^{\eta} \qsubrm{k}{eff}(\tilde{\eta})d\tilde{\eta}} 
\nonumber \\
&+& \frac{c_2}{\sqrt{\qsubrm{k}{eff}(\eta)}} e^{-i \int^{\eta} \qsubrm{k}{eff}(\tilde{\eta})d\tilde{\eta}}.
\label{WKBsolution}
\end{eqnarray}
Using the Wronskian condition (\ref{Wronskian}), we find $\left| c_2 \right|^2 - \left| c_1 \right|^2 = 1/2$ as 
a condition that the free parameters $c_1$ and $c_2$ have to fulfill. The most convenient choice is $c_1 = 0$ and $c_2 = 1/ \sqrt{2}$ which corresponds to a wave propagating in positive time direction. For this choice the mode function reads
\begin{equation}
v_k(\eta) = \frac{1}{\sqrt{2\qsubrm{k}{eff}(\eta)}} e^{-i \int^{\eta} \qsubrm{k}{eff}(\tilde{\eta})d\tilde{\eta}}.
\end{equation}
This represents a suitable choice because the mode function reduces to the Bunch-Davies vacuum in the 
UV limit ($k \rightarrow \infty$).\\
In order to check that this approach is valid, we plug this solution into Eq.~\eqref{osciequation}
and find that it is actually an exact solution to
\begin{equation}
v''_k + \left( \qsubrm{k}{eff}^2 - \frac{3}{4} \frac{(\qsubrm{k}{eff}')^2}{\qsubrm{k}{eff}^2} 
+ \frac{1}{2} \frac{{\qsubrm{k}{eff}''}}{\qsubrm{k}{eff}} \right) v_k = 0.
\end{equation}
Therefore, the solution \eqref{WKBsolution} is valid as long as
\begin{equation}
\left| \frac{1}{2} \frac{\qsubrm{k}{eff}''}{\qsubrm{k}{eff}^3} - \frac{3}{4} \frac{(\qsubrm{k}{eff}')
^2}{\qsubrm{k}{eff}^4} \right| \ll 1,
\label{WKBcondition}
\end{equation}
when the effective wavenumber, $\qsubrm{k}{eff}$, varies slowly. The appropriate initial conditions are then
\begin{align}
v_k(\qsubrm{\eta}{init}) &= \frac{1}{\sqrt{2 \qsubrm{k}{eff}(\qsubrm{\eta}{init})}}, \\
\left.\frac{d v_k}{d\eta} \right|_{\eta = \qsubrm{\eta}{init}} &= - \left( i \qsubrm{k}{eff} + 
\frac{1}{2} \frac{\qsubrm{k}{eff}'}{\qsubrm{k}{eff}} \right) \left.\frac{1}{\sqrt{2 \qsubrm{k}{eff}}} \right|_{\eta = \qsubrm{\eta}{init}},
\label{WKBinitialcond2}
\end{align}
where the exponential term can be neglected because it contributes only with an arbitrary phase. The initial moment $\qsubrm{\eta}{init}$ has to be chosen such that the WKB conditions are satisfied at this particular moment of time for all modes $k$. By analyzing $\qsubrm{k}{eff}(\eta)$ and its time derivatives, we can find an appropriate $\qsubrm{\eta}{init}$ in the remote past. For the numerical computations, the choice of $\qsubrm{\eta}{init}$  is therefore arbitrary as long as the condition \eqref{WKBcondition} is fulfilled. The instantaneous vacuum can be used as well for setting initial conditions. The instantaneous vacuum choice relies on the minimal energy state of the system defined by the Hamiltonian. Therefore, the requirement
\begin{equation}
 \left. \frac{\qsubrm{k}{eff}'}{(2 \qsubrm{k}{eff})^{3/2}} \right|_{\eta = \qsubrm{\eta}{init}} = 0,
 \label{InstCondition}
\end{equation}
has to be satisfied (see Ref. \cite{vacuum} for instance). In fact, one can find that there exists $\qsubrm{\eta}{init}$ such that both conditions $\eqref{WKBcondition}$ and \eqref{InstCondition}, are fulfilled. For reasons of comparability of the two approaches, we use this choice. In such a case, any difference between the two approaches is due 
to the higher order contribution to \eqref{WKBinitialcond2}, which is present in case of 
the adiabatic vacuum-type normalization.\\
 The conditions for the validity of both the instantaneous and WKB vacua depend strongly on the evolution 
of the cosmological term $z''/z$ during the pre-bounce contracting phase. A direct calculation leads to
\begin{eqnarray}
\frac{z^{\prime\prime}}{z}&=&-a^2 \left( m^2-2H^2+2 \kappa m^2\frac{\bar{\phi}\dot{\bar{\phi}}}{H} 
+\frac{7}{2} \kappa \Omega \dot{\bar{\phi}}^2 \right.
\nonumber \\
&& \left. - \kappa^2 \Omega^2\frac{\dot{\bar{\phi}}^{4}}{2H^2}-3 \kappa \frac{\dot{\bar{\phi}}^4}{\qsubrm{\rho}{c}} \right).
\label{eq:zprimeprimez}
\end{eqnarray}
This expression is valid at all times. In order to  analyze the shape of the effective potential it 
is convenient to divide the evolution into three background phases as mentioned in Sec. III. Then, analytical approximations for every phase can be used respectively. During the pre-bounce classical contracting phase, when $\rho(t) \ll \qsubrm{\rho}{c}$, 
the scalar field undergoes an oscillatory behavior with an amplitude proportional to $\sqrt{\rho(t)}$. The Hubble parameter $H$ is proportional to $\sqrt{\rho}$ as well, whereas $\Omega \simeq 1$. Inserting these solutions into Eq.~\eqref{eq:zprimeprimez} yield terms which are proportional to different orders of $\sqrt{\rho}$. Averaging over the 
oscillatory contributions, which all have a characteristic oscillation time of $1/m$, gives
\begin{eqnarray}
 \left\langle \frac{z^{\prime\prime}}{z} \right\rangle &=& - a^2 \left( m^2 - \alpha_1 \sqrt{\kappa} 
 m \sqrt{\rho(t)} + \alpha_2 \kappa \rho(t) \right. \nonumber \\
&& \left.~~~~ +~ \alpha_3 \kappa \qsubrm{\rho}{c} \left( \frac{\rho(t)}{\qsubrm{\rho}{c}} \right)^2 \right),
\label{effpotPre}
\end{eqnarray}
where the constants $\alpha_i$ are determined by the averaged oscillations. Since the energy density is increasing for all times in the remote past, $\rho(t)$ becomes sufficiently small in the remote past. Thus the $m^2$-term will dominate for early times and therefore $z''/z \propto - m^2 a^2$. On a logarithmic scale as a function of $\ln (a/\qsubrm{a}{B})$ like in Fig. \ref{fig:effpot} the absolute value of the effective potential is then given by a straight line with gradient $-2$ and with a $\ln(|z''/z|)$-intercept of $2 \ln(\qsubrm{a}{B} m) = -27.26$. This result is obtained as well by a purely analytical analysis which is presented in Fig. \ref{fig:effpot}. We use that the Hubble parameter $H$ is approximated by $H(t) = \qsubrm{H}{0} (1+\frac{3}{2} \qsubrm{H}{0} t)^{-1}$ for the pre-bounce phase when neglecting the fast oscillations, where $\qsubrm{H}{0}$ denotes the initial Hubble parameter. $\qsubrm{H}{0}$ is determined by the mass and the parameter $\alpha$, namely 
$\qsubrm{H}{0} = - m/ 3 \alpha$. Integration leads to the analytical  solution of the scale factor in the pre-bounce phase
\begin{equation}
a(t) = a_{\ast} \left(2 - \frac{m}{\alpha} t \right)^{\frac{2}{3}},
\end{equation}
where the prefactor $a_{\ast}$ is the scale factor for $t = \alpha/m$. With this expression the analytical solution for $\ln \left| (z''/z) \right| = \ln a(t)^2 m^2$ reads on the logarithmic scale
\begin{equation}
\ln \left|\frac{z''}{z}\right| = 2 \ln a_{\ast} + \frac{4}{3} \ln \left(2 - \frac{m}{\alpha} t \right) + 2 \ln m.
\end{equation}
As a function of $\ln (a /\qsubrm{a}{B} )$, one gets the red line on the left in Fig.~\ref{fig:effpot}. This analytic solution is valid until the energy density starts to dominate over the constant mass term in Eq.~\eqref{effpotPre}. The first term which is comparable to $m^2$ is proportional to $\sqrt{\rho}$. 
With the analytic solution of $\sqrt{\rho}$ 
in the pre-bounce phase,
\begin{equation}
\sqrt{\frac{\rho(t)}{\qsubrm{\rho}{c}}} = \frac{\Gamma \alpha^{-1}}{1 - 
\frac{1}{2\alpha} \left[ m t + \frac{1}{2} \sin (2mt + 2 \qsubrm{\theta}{0}) \right]},
\label{rhoanalytic}
\end{equation}
we can compute the time when the $\sqrt{\rho}$-term crosses `$m^2$' in its amplitude. 
The oscillation term in Eq. \eqref{rhoanalytic} is averaged over $T= 1/m$. This transition point is referred to 
$\qsubrm{N}{pre} := \ln (\qsubrm{a}{pre} /\qsubrm{a}{B}) = - 5.38$ in the figure.\\
During the bouncing phase the potential energy parameter $x$, is very small compared to the kinetic potential parameter $y$, since we consider a kinetic bounce scenario. Then, in particular, $x^2 \ll y^2$ is satisfied during this phase and the Hubble parameter can be reduced to
\begin{equation}
H^2  \simeq \frac{\kappa \qsubrm{\rho}{c}}{3} y^2 \left( 1 - y^2 \right).
\end{equation}
The analytic solution for $y$ around the bounce is given by $y(t) = (1 + 3 \kappa \qsubrm{\rho}{c} (t-\qsubrm{t}{B})^2)^{-1/2}$, as presented in Ref.~\cite{boris} and the scale factor is related to $y$ via $a = \qsubrm{a}{B} \left| y \right|^{-1/3}$. With these 
approximations the expression for $z''/z$ reduces to
\begin{equation*}
\frac{z''}{z} = \frac{\kappa \qsubrm{\rho}{c} \qsubrm{a}{B}^2}{3} 
\frac{\left(-\left( \frac{a}{\qsubrm{a}{B}}\right)^2 + 23 \left( \frac{a}{\qsubrm{a}{B}}\right)^{-4} 
- 4 \left( \frac{a}{\qsubrm{a}{B}}\right)^{-10} \right) }{\left( \frac{a}{\qsubrm{a}{B}}\right)^6 - 1}.
\end{equation*}
Note that this expression is positively valued and diverges at the bounce. The absolute 
value of this expression on the logarithmic scale and as a function of $\ln (a/ \qsubrm{a}{B})$ provides 
the red lines around the bounce in Fig.~\ref{fig:effpot}. These approximations are valid until $x^2$ 
becomes significant in comparison to $y^2$, let's say $x^2 > y^2/10$. The analytic solutions for $x$ 
and $y$ provide these transition points of validity respectively before and after the bounce, namely 
$\qsubrm{N}{preB} := \ln (\qsubrm{a}{preB}/ \qsubrm{a}{B}) = - 4.41$ and  $\qsubrm{N}{postB} := 
\ln (\qsubrm{a}{postB}/ \qsubrm{a}{B}) = 3.62$, as shown in Fig.~\ref{fig:effpot}.\\
During slow-roll inflation, $\Omega\simeq1$ such that $z''/z$ takes its classical expression. This leads to $z''/z=(2+6\qsubrm{\epsilon}{H}-3\qsubrm{\delta}{H})/\eta^2$, with $\qsubrm{\epsilon}{H}$ and $\qsubrm{\delta}{H}$ the first and second Hubble flow functions, both much smaller than unity during inflation. Furthermore, $a \propto 1/ \eta$ and therefore 
$\ln(z''/z)\propto2\ln(a/\qsubrm{a}{B})$, see Fig.~\ref{fig:effpot}. In particular the effective potential 
is then given by $z''/z = \frac{1}{2} a^2 H^2$. The curve can be approximated from the beginning of 
slow-roll inflation, \textit{i.e.} when $\qsubrm{t}{i} = \qsubrm{t}{B} + f/m$ where $f$ is an analytical 
expression related to the Lambert function and $\qsubrm{t}{B}$ is given analytically as well. For this 
time $\qsubrm{a}{i} = \qsubrm{a}{B} \Gamma^{-\frac{1}{3}}$ and the logarithm of the absolute value 
of the effective potential is given by $\ln |z''/z| = \ln (2/\eta^2) = \ln ((1/2) \qsubrm{a}{i}^2 \qsubrm{H}{i}^2)$. 
The value of the Hubble parameter $\qsubrm{H}{i}$ is given analytically as well, see Ref.~\cite{boris}. 
The approximation is valid starting from $\qsubrm{N}{post} := \ln (\qsubrm{a}{post} / \qsubrm{a}{B}) = 5.53$. 
For the approximation we use that
\begin{equation}
H(t) = \qsubrm{H}{i} \left| 1 - \frac{\epsilon \Gamma}{\qsubrm{x}{i}} m (t - \qsubrm{t}{i}) \right|
\end{equation}
during slow-roll inflation, where $\epsilon$ is the sign of the cosine of the phase parameter between the
potential and kinetic energy parameters at the transition point between the pre-bounce and bouncing phases, and $\qsubrm{x}{i}$ is the value of the potential energy parameter at $\qsubrm{t}{i}$. Furthermore, the scale factor undergoes an exponential growth with coordinate time, namely
\begin{equation}
a(t) = \qsubrm{a}{i} e^{- \frac{\qsubrm{H}{i}}{2 \qsubrm{x}{i}} (t- \qsubrm{t}{i}) 
(\epsilon \Gamma m (t- \qsubrm{t}{i}) -2 \qsubrm{x}{i})}.
\label{scalefactorSlowroll}
\end{equation}
The analytic fit given by these two functions is displayed by the red line on the right side in 
Fig.~\ref{fig:effpot}. as a function of $\ln (a / \qsubrm{a}{B})$ with $a$ provided by 
Eq.~\eqref{scalefactorSlowroll}. The slight difference between numerical results and 
the analytical solution during slow-roll inflation is due to the fact that the analytical 
approximations for this phase goes back on approximations even for the pre-bounce 
phase. Hence, small differences are propagated.
\section{The scalar power spectrum}
For scalar modes, the primordial power spectrum at the end of inflation is defined in terms of the mode functions by virtue of the definition \eqref{PowerSpectrum}. As we shall see in the next section, three ranges of wavenumbers can be identified, depending on how they compare to the effective potential $z''/z$: (i) the infrared regime, 
(ii) intermediate scales and (iii) the ultraviolet regime.\subsection{The infrared regime}
The infrared limit (IR) of the primordial power spectrum corresponds to modes such that 
$k^2 \ll \left| z''/z \right| $ during the pre-bounce contracting phase. These modes are frozen 
during the bouncing phase and slow-roll inflation. The transition between the contracting phase 
and the bouncing phase occurs when $H \simeq - m/3$, as discussed in Ref.~\cite{boris}. At the 
transition, the effective potential term $z''/z$ is well approximated by $a''/a$, since $\dot{\bar{\phi}}/H$ 
remains nearly constant. This allows us to introduce the following IR scale (see Ref.~\cite{boris}):
\begin{equation}
\qsubrm{k}{IR} := \frac{\qsubrm{a}{B}}{3 \sqrt{2}} \left( m^2\sqrt{3\kappa\qsubrm{\rho}{c}}\right)^{1/3}
\approx 4.7 \times 10^{-5} \qsubrm{m}{Pl},
\end{equation}
where the numerical value has been obtained for $m = 1.2 \times 10^{-6} \qsubrm{m}{Pl}$, 
$\qsubrm{a}{B} = 1$ and $\qsubrm{\rho}{c} = 0.41 \qsubrm{m}{Pl}^4$. \\
There is an important difference with respect to the case of  tensor modes. For small enough values 
of $k$ in the IR regime, the tensor power spectrum tends to be scale-invariant. This is due to the 
fact that initial conditions for  tensor modes are set when \textit{all} the modes of interest are 
sub-Hubble (or, more precisely, $k^2 \gg a''/a$). For the scalar modes, however, it is impossible to 
set appropriate initial conditions at a time when all relevant modes are such that $k^2 \gg \left| z''/z \right|$. 
Indeed,  the conditions discussed in the previous section, \eqref{WKBcondition} and \eqref{InstCondition}, have to be satisfied as well respectively for the WKB and the instantaneous vacuum-type normalizations. In addition, the absolute value of the effective potential term keeps decreasing in the remote past of the contracting branch. It is possible to find a time $\qsubrm{\eta}{init}$, in the classical contracting phase, at which (i) the absolute value of the 
effective potential term $z''/z$ is close to zero and (ii) the conditions of validity of the vacuum states 
are fulfilled. Nevertheless for the WKB vacuum, at this particular time, when $\left| z''/z \right|$ is minimal and condition \eqref{WKBcondition} is satisfied, the effective potential does not strictly vanish, it is $\left| z''/z \right|=2.1 \times 10^{-7} \qsubrm{m}{Pl}$, and therefore only the modes with $k>4.5 \times 10^{-4} \qsubrm{m}{Pl}$ satisfy the condition $k^2 > \left| z''/z \right|$.\subsection{The ultraviolet regime}
In the deformed algebra approach, the ultraviolet modes (UV) experience an exponential growth with 
increasing wavenumbers. This is due to the $\Omega$-factor in front of the wavenumber in Eq.~\eqref{eomR}, which becomes negative near the bounce. When approaching the bounce, the friction term in Eq.~\eqref{eq:EoMBounce}, namely $\dot{H}/H=1/(t-\qsubrm{t}{B})$, diverges. However, the 
approximate solution given in section IV shows that $\dot{\mathcal{R}}_k$ vanishes at the bounce, 
since its generic expression is given by
\begin{equation}
\dot{\mathcal{R}}_k = (t- \qsubrm{t}{B}) \left[ c_1 \text{ch} 
\left( k (t- \qsubrm{t}{B}) \right) - c_2 \text{sh} \left( k ( t- \qsubrm{t}{B}) \right) \right],
\end{equation}
where $c_1$ and $c_2$ are numerical constants that have to be chosen in accordance to the initial state of the perturbations. Thus, the equation of motion \eqref{eq:EoMBounce} has no singularity, and $(\dot{H}/H)\dot{\mathcal{R}}_k \propto\sqrt{k}$. We are left with the differential equation of an harmonic oscillator, but with an imaginary frequency and a constant term, say $\beta \sqrt{k}$, such that $\ddot{\mathcal{R}}_k + \beta \sqrt{k }- k^2 \mathcal{R}_k = 0$. Close to the bounce the generic solution to this equation is 
\begin{equation*}
\mathcal{R}_k =\beta k^{-\frac{3}{2}}+ \alpha_+ e^{kt} + \alpha_- e^{-kt}.
\end{equation*}
So, in the large $k$ limit and close to the bounce the amplitude of  scalar modes receives 
a real exponential contribution, which marks the Euclidean nature of the bounce ($\Omega<0$). For large scales however, one has $\left|\Omega k^2\right|\ll\left|z''/z\right|$ around the bounce. The solutions for the mode functions are $\mathcal{R}_k(\eta)\sim A_k+B_k\int^\eta d\eta'/z^2(\eta')$, and the large-scale modes are thus qualitatively not affected by the Euclidean nature of the bounce. A similar behavior for the tensor modes was already discussed in Refs.
\cite{linda_pk} and \cite{boris}.\\
The characteristic energy scale $\qsubrm{k}{UV}$  beyond which the effect of the Euclidean 
nature of the bounce qualitatively affects the evolution of the modes can be determined from an analysis 
of $\qsubrm{k}{eff}^2$  in Eq.~\eqref{osciequation}. In vicinity of the bounce $\Omega \approx -1$ and $z''/z >0$. Therefore, a given mode has an imaginary time-dependent wavenumber for a certain period around the bounce, {\it i.e.} `$\qsubrm{k}{eff}^2 <0$'. This is what we call the `Euclidean phase' in this approach. However, the imaginary effective wavenumber only plays a role, if the interval of conformal time spend by the mode in this regime is large enough. Of course, in the Euclidean phase, it makes physically no sense to talk about time, the evolution parameter $\eta$ however remains and quantifies the `period' of the mode. If the mode spends more than one period in the region with complex effective wavenumber, $\qsubrm{k}{eff}^2 <0$, the mode will be amplified significantly. This is in accordance with the analytical solutions to the approximated differential equation around the bounce, \eqref{eq:EoMBounce}, which show an hyperbolic behavior for $k(t-t_B)\gg1$. 
This leads to the following condition for the energy scale $\qsubrm{k}{UV}$: 
\begin{equation}
\qsubrm{k}{UV} \Delta \eta(\qsubrm{k}{UV}) \approx 1.
\end{equation}
A direct analytical analysis of this condition gives the following expression 
\begin{equation}
\qsubrm{k}{UV}  \simeq \qsubrm{a}{B} \sqrt{\frac{2}{3}}\sqrt{\frac{\sqrt{2}}{2\sqrt{2}-1}}
\sqrt{\kappa \qsubrm{\rho}{c}} \approx 2.3\ \qsubrm{m}{Pl}, 
\end{equation}
where the numerical value is obtained with use of $\qsubrm{a}{B} = 1$ and $\qsubrm{\rho}{c} = 0.41\ \qsubrm{m}{Pl}^4$.

\subsection{Numerical Results}
\begin{figure}[H]
\includegraphics[scale=0.41]{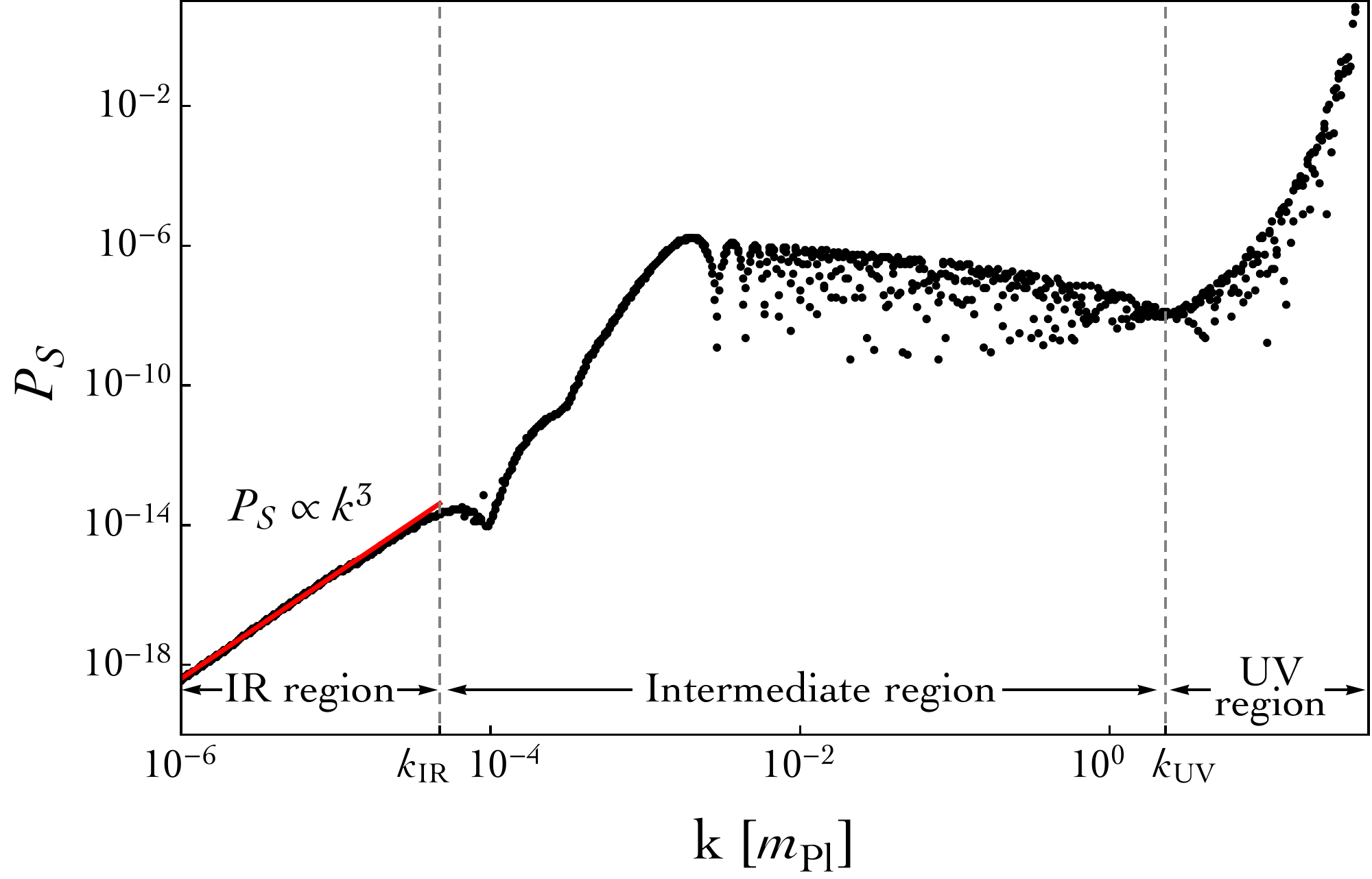}
\caption{Primordial power spectrum for scalar modes in the deformed algebra approach for 
$m=1.2 \times 10^{-6} \qsubrm{m}{Pl}$ and the adiabatic vacuum for initial conditions in the 
pre-bounce (classical) contracting phase. The cosmological background is fixed such that 
$\qsubrm{x}{B}= 3.55 \times 10^{-6}$ and $\qsubrm{a}{B}=1$ at the bounce.}
\label{fig:wkbpk}
\end{figure}

\begin{figure}[H]
\includegraphics[scale=0.41]{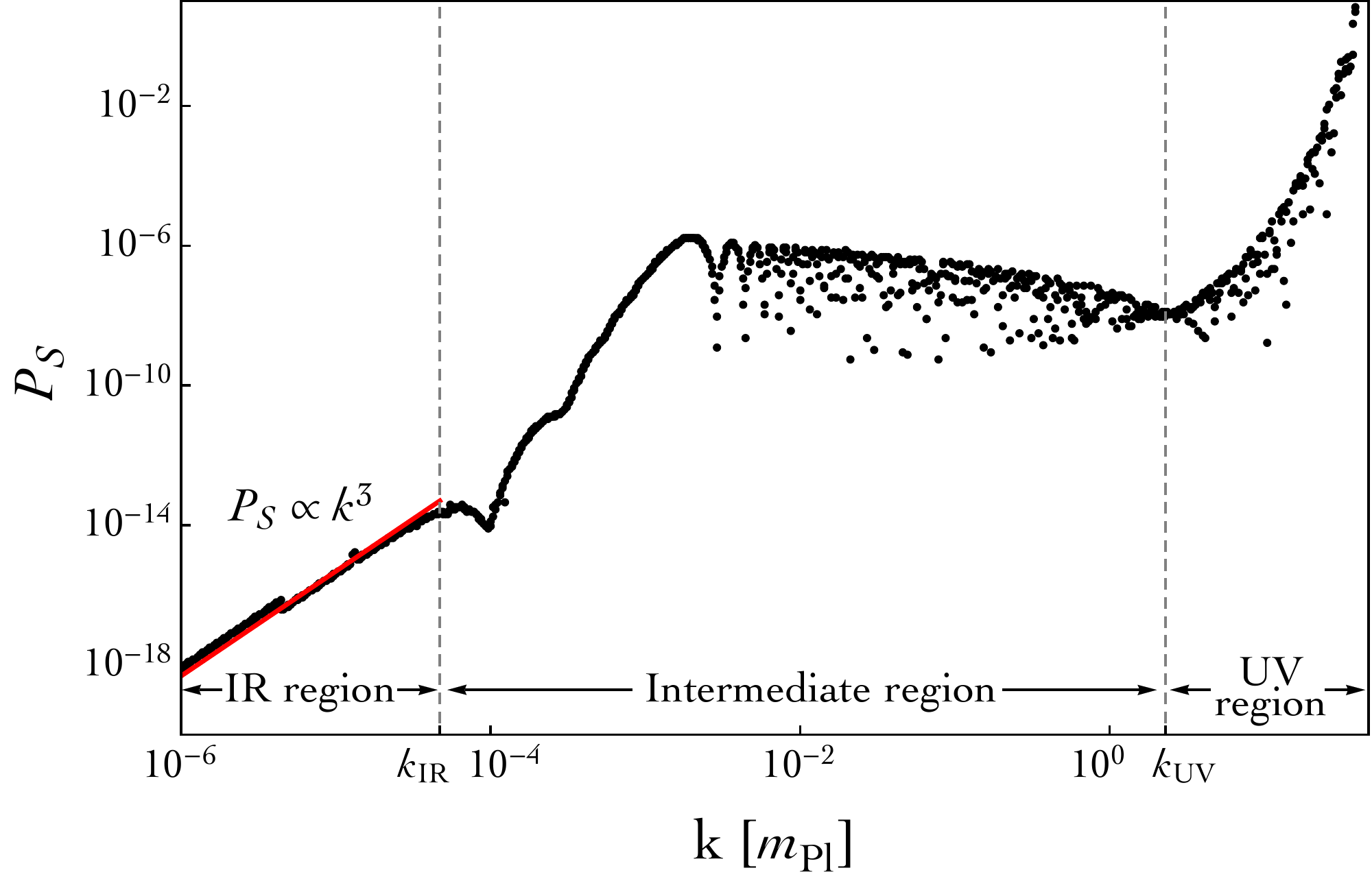}
\caption{Primordial power spectrum for scalar modes in the deformed algebra approach 
for $m=1.2 \times 10^{-6} \qsubrm{m}{Pl}$ and the instantaneous vacuum for initial 
conditions in the pre-bounce (classical) contracting phase. The cosmological background is 
fixed such that $\qsubrm{x}{B}= 3.55 \times 10^{-6}$ and $\qsubrm{a}{B}=1$ at the bounce.}
\label{fig:instpk}
\end{figure}

The scalar power spectrum is obtained by numerical integration of the equation of motion for the mode functions, respectively for the variable $h$ and $\mathcal{R}$ for different phases in the time evolution, and the solution of the background equations \eqref{eq:sys}. Initial conditions for the perturbations can be set according to the WKB approximation 
referring to the adiabatic vacuum, or with the instantaneous vacuum as shown in Sec. V. The initial conditions for the cosmological background are set in the contracting phase, such that the preferred value of the potential energy parameter $x$ at the bounce is obtained. Note that the dynamics of the background and subsequently the shape of the power spectrum 
$\qsubrm{\mathcal{P}}{S}(k)$ are determined by the mass $m$ of the scalar field, the value of the critical energy density $\qsubrm{\rho}{c}$ and the phase $\theta_0$. \\
Our numerical results are shown in Fig.~\ref{fig:wkbpk} and Fig.~\ref{fig:instpk} which display the primordial 
power spectra for the scalar modes, choosing the adiabatic (WKB) vacuum as initial conditions (Fig.~\ref{fig:wkbpk}) 
and the instantaneous vacuum as initial conditions (Fig.~\ref{fig:instpk}). The three regions mentioned in 
the previous section ($k<\qsubrm{k}{IR}, \qsubrm{k}{IR}<k<\qsubrm{k}{UV}, k>\qsubrm{k}{UV}$) can be well identified in the spectra. \\
In the intermediate region ($\qsubrm{k}{IR}<k<\qsubrm{k}{UV}$), the spectrum follows a characteristic oscillating behavior observed also in case of the tensor modes (see Ref.~\cite{boris}). For the 
values of $k>\qsubrm{k}{UV}$, the power spectrum is characterized by the exponential growth. 
This behavior should, however, be considered with care. First, the UV regime ($k>\qsubrm{k}{UV}$) corresponds to the modes which are trans-Planckian at the bounce. For such modes the effective description based on the continuous equations of motion might not be reliable. Second, the observed amplification is due to an instability related to the elliptic type of the equation of motion for perturbations in the Euclidean regime. The Cauchy initial value problem might not be valid for the modes with $k>\qsubrm{k}{UV}$ which are strongly affected by the Euclidean nature of the deep quantum regime.
The spectra for the adiabatic vacuum and the instantaneous vacuum choices are almost
identical. The only difference is a slight enhancement in the IR region for the instantaneous vacuum-type 
normalization, in comparison to the adiabatic vacuum choice. This effect is due to the difference 
in $v'_k(\eta_{\text{init}})$ for both types of initial conditions. At values of $k<\qsubrm{k}{IR}$, the 
shape of the spectra is mostly due to the initial super-Hubble vacuum normalization. 
In this limit $\qsubrm{k}{eff}\simeq \sqrt{k^2 - \frac{z''}{z}} \approx$ const, is almost the same for 
every mode at this particular initial time, because $k^2\ll \left|\frac{z''}{z}\right|$. 
Therefore, $\qsubrm{\mathcal{P}}{S}(k) =  \frac{k^3}{2\pi^2} \frac{1}{2\qsubrm{k}{eff} z^2} \propto k^3$. 
The initial $\qsubrm{\mathcal{P}}{S}(k) \propto k^3$ behavior is preserved in the further evolution 
due to the super-Hubble nature of the modes.  Only the absolute amplitude changes, which is a result of the 
time dependence of the $z$ parameter.\\
Furthermore, it is worth stressing that at  scales $k<\qsubrm{k}{UV}$ the power spectra computed in this paper qualitatively agree with those obtained in the so-called ``dressed metric" approach to perturbations in LQC \cite{Agullo:2015tca}. In that case the $\Omega$-factor does not appear in front of the Laplace operator and the instabilities related to the Euclidean phase 
do not arise. Therefore, the corresponding spectrum at $k>\qsubrm{k}{UV}$ becomes nearly-scale 
invariant in the ``dressed metric" approach,  as in the standard inflationary picture.

We live the detailed associated phenomenology for a future study \cite{next_boris}. It is however important to stress that the spectrum derived in this article is basically in disagreement with data that are very precise for scalar modes. This rules out neither LQC in general nor the {\it deformed algebra} as a whole. But this shows that the set of specific hypotheses presented here is obviously in tension with measurements. Ruling out a given setting is useful for future quantum gravity investigations.

\section{Discussion and conclusion}
In this work, the primordial scalar power spectrum in the so-called ``deformed algebra" approach to perturbations in loop quantum cosmology has been derived. Our considerations were focused on the model with a massive scalar field. The instantaneous and adiabatic vacuum-type initial conditions were imposed in the contracting phase. The non-trivial issue in the evolution of modes is their behavior in the Euclidean phase ($\Omega<0$) surrounding the bounce. In this region, the equation of motion for the mode functions changes its type from hyperbolic to elliptic. In such a case, no preferred time direction exists. 
It is usually argued that a signature change implies instabilities because the oscillating time dependence `$\exp(\pm i\omega t)$' is replaced by an exponential `$\exp(\pm \omega t)$' one. The growing mode leads to an instability if initial values are chosen at some fixed $t$. The approach suggested in Ref.~\cite{silent2} precisely investigates how a boundary value problem for this kind of elliptic equations can eliminate the instability. This is, certainly, a path worth investigating. This article is devoted to the other hypothesis: considering seriously this real argument in the exponential function and investigate its physical consequences.\\
Even if the problem is not posed in the usual way for the partial differential equation in that case, reliable predictions can still be obtained for sufficiently low values of $k$ in the Fourier space representation. More precisely, a characteristic scale $\qsubrm{k}{UV}$ discriminates between the modes which are amplified due to the imaginary effective wavenumber $\qsubrm{k}{eff}$ around the bounce. This is a mathematical consequence of the equation of motion in Fourier space. Of course, the physical interpretation as having or not enough ``time" to oscillate and feel the quantum geometrical structure near the bounce is not possible without time. But this is not something fundamentally new or surprising in quantum cosmology/gravity. For the modes 
satisfying $k>\qsubrm{k}{UV}$, the elliptic nature of the equations becomes important, leading to an abnormal amplification of the power spectrum. The effect is the same as the one observed earlier in case of the tensor perturbations \cite{linda_pk}. In turn, for $k<\qsubrm{k}{UV}$, that is for large wavelengths, the modes are not subject for a sufficiently long period to the negative effective potential, $\qsubrm{k}{eff}^2 <0$. Thus, the corresponding modes are not affected by the hyperbolic amplification, as discussed above. In the regime, $\qsubrm{k}{IR}<k<\qsubrm{k}{UV}$, a typical oscillatory behavior is observed. In the IR limit, the shape of the 
spectrum is determined by the initial vacuum normalization and scales as $\qsubrm{\mathcal{P}}{S}(k) \propto k^3$. This behavior is very different from the one observed in case of the tensor modes (see Ref.~\cite{boris}), where the power spectrum becomes nearly-scale invariant while $k\rightarrow 0$. This is 
because the massive scalar field, oscillating in the contracting branch, effectively behaves as dust 
matter. As it is known, the freezing of massless modes during such an evolution leads to scale-invariance 
of the power spectrum, as for the case of tensor perturbations. For the scalar perturbations in a 
model with a massive scalar field the gauge-invariant degree of freedom $\qsubrm{v}{S}$ is explicitly 
massive leading to a breakdown of the scale-invariance.\\
Several points of the picture presented in this study still need to be addressed. First, the observational 
consequences of this calculations should be studied into the details. The key point for phenomenology is the knowledge of full duration of inflation. This is what translates coordinate wavenumbers used in this study into physical wavenumbers that can be compared with data.  The number of inflationnary e-folds is in one-to-one correspondence with the value of the scalar field at the bounce which, itself, depends on the phase of the oscillations in the contracting branch. The higher the field value (and therefore the fraction of potential energy) at the bounce, the longer the inflation period and the smaller the physical scales at the bounce that are nowadays probed by astronomical observations. We leave a detailed phenomenological study for a future work, \cite{next_boris}, but it is easy to guess from the primordial power spectrum derived in this study that the model as it is here investigated disagrees with data. It could simply be a consequence of the way modes with $k>1$ are handled. But whatever the reason this problem should be stressed. Second, other 
proposals for setting initial conditions should also be considered. Here, the subtle issue of the very 
meaning of time in the Euclidean phase were deliberately ignored: modes were naively propagated 
through the Euclidean phase. A proper addressing of the well-posedness is crucial to obtain stable solutions in the $k>\qsubrm{k}{UV}$ regime (even if their physical meaning is not clear due to the breakdown of validity of the effective equations under considerations) \cite{next_jakub}. Furthermore, the matter content considered in this paper is no more favored by the 
observations of the cosmic microwave background radiation. A caraful analysis of different inflationary potentials would therefore be desirable. In particular, the Colemen-Weinberg  potential with an unstable state may lead to inflationary spectra being in agreement with the up-to-date observational data. Such a change
of the potential function would unavoidably affect our predictions regarding the shape of the power 
spectra.
\acknowledgments
BB is supported by a grant from ENS Lyon. JM is supported by the Grant 
DEC-2014/13/D/ST2/01895 of the Polish National Centre of Science.

\end{document}